\documentclass[10pt,tightenlines,aps,prd,twocolumn,showpacs,preprintnumbers,amsmath,amssymb]{revtex4}
\usepackage{graphicx,color}
\definecolor{darkblue}{rgb}{0,0,0.7}
\definecolor{darkred}{rgb}{0.7,0,0}
\usepackage[dvips, colorlinks, citecolor=darkblue, linkcolor=darkred, urlcolor=blue]{hyperref}
\allowdisplaybreaks
\begin{document}

\title{Displacement-noise-free resonant speed meter for gravitational-wave detection}
\author{Sergey P. Vyatchanin}
\affiliation{Faculty of Physics, Moscow State University, Moscow,
119992, Russia} \email{svyatchanin@phys.msu.ru}
\date{\today}

\begin{abstract}
We demonstrate that speedmeter, based on double pumped resonant Sagnac interferometer, can be used as a displacement noise free gravitational-wave (GW) detector. The displacement noise of cavity mirrors can be completely excluded through a proper linear combination of the output signals. We show that in low-frequency region the obtained displacement-noise-free response signal is stronger than the one in previously proposed displacement noise free interferometers.
\end{abstract}

\maketitle

\section{Introduction}\label{sec_intro}
Currently there is international ``community'' of first generation laser interferometric gravitational wave (GW) detectors \cite{1972_EMW_detector,2000_GW_detection} (LIGO in
USA \cite{1992_LIGO,2006_LIGO_status,website_LIGO}, VIRGO in Italy
\cite{2006_VIRGO_status,website_VIRGO}, GEO-600 in Germany
\cite{2006_GEO-600_status,website_GEO-600}, TAMA-300 in Japan
\cite{2005_TAMA-300_status,website_TAMA-300} and ACIGA in Australia
\cite{2006_ACIGA_status,website_ACIGA}). The development of the
second-generation GW detectors (Advanced LIGO in USA
\cite{2002_Adv_LIGO_config,website_Adv_LIGO}, LCGT in Japan
\cite{2006_LCGT_status}) is underway.
The ultimate sensitivity of laser GW detectors is restricted by the Standard Quantum Limit (SQL) --- a specific sensitivity level where the measurement noise of the meter (photon shot noise) is equal to its back-action (radiation pressure noise) \cite{1968_SQL,1975_SQL,1977_SQL,1992_quant_meas}. The sensitivity of GW detectors is also limited by classical displacements noises of various nature: seismic and
gravity-gradient noise at low frequencies (below $\sim 50$ Hz), thermal noise in suspensions, bulks and coatings of the mirrors ($\sim 50\div 500$ Hz).

In 2004 S. Kawamura and Y. Chen put forward an idea of so called displacement-noise-free
interferometer (DFI) which is free from displacement noise of the test masses as well as from optical laser noise \cite{2004_DNF_GW_detection, 2006_DTNF_GW_detection,2006_interferometers_DNF_GW_detection}.
The most attractive feature of DFI is the straightforward overcoming of  SQL (since the radiation pressure noise is canceled) without the need of implementing complicated schemes for Quantum-Non-Demolition (QND) measurements \cite{1981_squeezed_light,1982_squeezed_light,2002_conversion,1996_QND_toys_tools}.

The possibility of GW signal separation from displacement noise of the test masses is based on the the {\em distributed} interaction of GW with light wave in contrast with localized influence of mirrors positions on the light wave only at the moments of reflection. The ``price'' of this separation is decrease in GW response, which is more obvious at low frequencies --- so called long wave approximation when the distance $L$ separating test masses is much less than the gravitational wave length $\lambda_\text{gw}$, i.e. $L/\lambda_\text{gw}\ll 1$ or $\Omega_{\textrm{gw}}\tau\ll 1$ ($\tau =L/c$ is time of light trip between test masses, $c$ is light speed and $\Omega_\text{gw}=2\pi\, c/\lambda_\text{gw}$ is mean frequency of GW). In particular, the analysis presented in \cite{2006_interferometers_DNF_GW_detection} for double Mach-Zehnder interferometer showed that in long wave approximation the shot-noise limited sensitivity to GWs turns out to be limited by $(\Omega_{\textrm{gw}}\tau)^2$-factor for 3D configurations and $(\Omega_{\textrm{gw}}\tau)^3$-factor for 2D configurations. For a signals about  $\Omega_{\textrm{gw}}/2\pi\approx 100$ Hz and $L\approx 4$ km ($\tau\simeq 10^{-5}$~s), DFI sensitivity of the ground-based detector is $(\Omega_\text{gw}\tau)^3\simeq 10^{-6}$ times worse than that of a conventional single round-trip laser detector. 

Another approach to displacement noise cancellation was presented in \cite{2008_toy} where a single detuned Fabry-Perot cavity pumped through both movable, partially transparent mirrors was analyzed. Two double pumped Fabry-Perot cavities positioned in line (1D configuration) represent  DFI with decrease of GW response by $(\Omega_{\textrm{gw}}\tau)^2$-factor only \cite{08a1RaVy}.

Recently N.~Nishizawa, S.~Kawamura and M.~Sakagami proposed using resonant speed meter based on resonant Sagnac interferometer as DFI for GW detection \cite{2008Kawamura}. In particular, the authors  demonstrated that displacement noise cancellation is possible, albeit  within a narrow band.

In this paper we investigate the model originated from a model of DFI speed meter \cite{2008Kawamura}, design and analysis are presented in Sec.~\ref{simple}. In Sec.~\ref{DFI} we propose and analyze double pumped resonant speed meter --- it gives the possibility to exclude information on displacement and laser noise completely over a {\em wide} spectral range. The ``price'' for isolation of the GW signal from displacement noise deals with suppression of sensitivity by a relatively modest factor of $(\Omega_{\textrm{gw}}\tau)$ as compared with conventional interferometers --- it is much larger than the limiting factor $(\Omega_{\textrm{gw}}\tau)^3$ for  double Mach-Zehnder 2D configuration \cite{2006_interferometers_DNF_GW_detection} or $(\Omega_{\textrm{gw}}\tau)^2$ for two double pumped Fabry-Perot cavities \cite{08a1RaVy}.

\section{Speed meter based on Sagnac interferometer}\label{simple}

\begin{figure}
\includegraphics[width=0.45\textwidth]{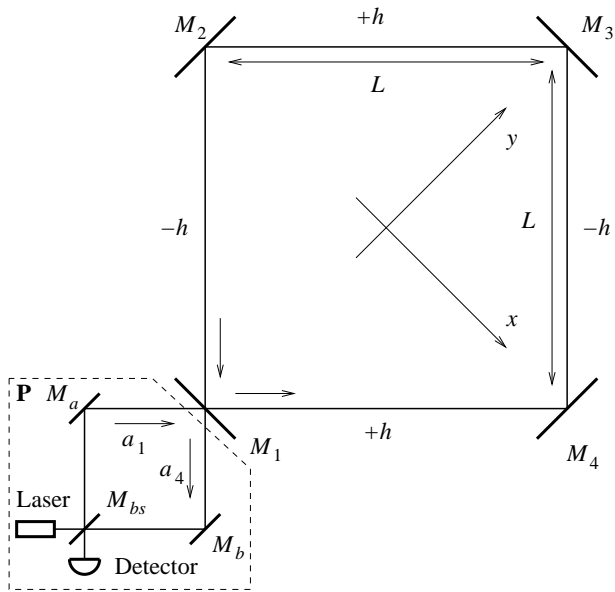}
\caption{Simplified model of resonant speed meter based on Sagnac interferometer. Laser, detector and auxiliary mirrors $M_{bs},\ M_a,\ M_b$ are rigidly mounted on platform $P$. GW propagates perpendicularly  to plane of interferometer. Recycling cavity is composed of mirrors $M_1,\ M_2,\ M_3,\ M_4$. Light waves circulating clockwise are shown by arrows.}\label{sd}
\end{figure}

For clear demonstration we start with analysis of the simplest model of speed meter based on single pumped resonant Sagnac interferometer \cite{2008Kawamura} shown in Fig.~\ref{sd}. It differs from conventional Sagnac interferometer by having an additional resonant ring cavity. For simplicity we assume that neither mirror has optical losses. Laser, detector, beam splitter $M_{bs}$ with 50\% transmissivity and completely reflective mirrors $M_a,\ M_b$ are rigidly mounted on platform $P$ and do not move relatively it. Platform $P$ can move as a free mass along axis $y$ (the coordinate frame is shown in Fig.~\ref{sd}). Laser beams, divided by beam splitter $M_{bs}$, are reflected by completely reflective mirrors $M_a$ and $M_b$ and enter the ring shaped cavity through input mirror $M_1$. Ring cavity is formed by input mirror $M_1$ with small amplitude transmissivity $T$ ($T\ll 1$) and completely reflective mirrors $M_2,\ M_3,\ M_4$. In cavity each beam circulates clockwise or counterclockwise, then they leave the cavity and finally recombine at the beam splitter $M_{bs}$. Mirrors $M_1,\ M_2,\ M_3.\ M_4$ can move as free masses. We consider only displacements of mirrors corresponding to deformation of ring cavity, i.e. displacements $y_1,\ y_3$ of mirrors $M_1,\ M_3$, displacement $x_2,\ x_4$ of mirrors $M_2,\ M_4$ and displacement $y_P$ of platform $P$ because of fluctuations of these displacements mask GW signal. We do not consider rotation of interferometer as a whole. Our aim is to exclude fluctuational displacements $y_1,\ x_2,\ y_3,\ x_4,\ y_P$.

Without the loss of generality we assume the interferometer to be lying in the plane perpendicular to direction of GW and directions of light propagations coincide with the GW principal axes. Below for wave propagating between mirrors $M_1$ and $M_4$ (or between $M_2$ and $M_3$) we put dimensionless metric with positive sign ($+h(t)$) and with negative sign ($-h(t)$) in normal direction as shown in Fig.~\ref{sd}. We assume that dimension of platform $P$ is relatively small, neglecting GW effect on light propagating between mirrors $M_{bs},\ M_a,\ M_b$ and taking into account only GW influence on wave propagating inside the ring cavity. Due to small size of platform $P$ we also assume that the phase advance of light propagating between mirrors of platform does not depend on frequency.

Note that radiation emitted from the laser is registered (after circulation in interferometer) by detector on the same platform $P$. Actually detector is a homodyne detector measuring the quadrature amplitude of the output wave (the local oscillator wave is the one slitted from the same laser). Strictly speaking, in order to describe detection of light wave we have to work in the reference frame of detector, i.e. in accelerated frame. However, in our model the detector is mounted on the same platform as the laser whose radiation is registered by detector and, hence, we can work in inertial laboratory frame as it was demonstrated in \cite{2008_toy,2008_Tarabrin}. Moreover, in this case of round trip configuration we can use transverse-traceless (TT) gauge considering GW action as effective modulation of refractive index $(1+h(t)/2)$ by weak GW perturbation metric $h(t)$. It is worth noting that in the opposite case, when laser and detector are mounted on different platforms, we should use the local Lorentz (LL) gauge --- see details in \cite{2008_Tarabrin}.

We assume that interferometer is tuned so that for the case when mirrors and platform are at rest and GW is absent the light from the laser after circulating inside the interferometer returns to the laser port in a similar manner as the vacuum fluctuations wave from detector return to detector port. Small perturbations of mirrors positions or GW action produce appearance of the light in detector port. Similar regime of dark port is planned to use in Advanced LIGO. This regime allows excluding laser noise, because in detector port experimenter measures only signal (providing information on displacement and GW), and vacuum fluctuations pumped into interferometer through detector port.

It is convenient to represent the electric field operator $E(x,t)$ of the light wave emitted by laser and propagating, for example, along $z$-axis as a sum of ``strong'' (classical) monochromatic wave (which approximates the light beam having cross-section $S$) with amplitude $A$, frequency $\omega$ ($k=\omega/c$ is wave vector, $c$ is light speed) and ``weak'' wave describing quantum fluctuations of the electromagnetic field:
\begin{subequations}
\label{E}
\begin{align}
E(x,t)&=\sqrt{\frac{2\pi\hbar\omega}{Sc}}\,
\Bigl[A+a(x,t)\Bigr]e^{-i(\omega t-kz)}+{\textrm{h.c.}},\\
a(x,t)&=\int_{-\infty}^{+\infty}
a(\omega+\Omega)e^{-i\Omega\left(t- z/c\right)}\,\frac{d\Omega}{2\pi},
\end{align}
\end{subequations}
with amplitude $a(\omega+\Omega)$ (Heisenberg operator to be
strict) obeying the commutation relations:
\begin{align*}
\bigl[a(\omega+\Omega),a(\omega+\Omega')\bigr]&=0,\\
\bigl[a(\omega+\Omega),a^+(\omega+\Omega')\bigr]&=2\pi\delta(\Omega-\Omega').
\end{align*}
This notation for quantum fluctuations $a(\omega+\Omega)$ is convenient since it coincides exactly with Fourier representation of classical fields. Below we denote $a(\Omega)\equiv a(\omega+\Omega)$ and we omit the $\sqrt{2\pi\hbar\omega2/Sc}$-multiplier. For convenience throughout the paper we denote mean amplitudes by block letters and corresponding small additions by {\em the same} small letter as in (\ref{E}). We assume that input laser wave is in coherent state, it means that fluctuational amplitude $a(\Omega)$ as well as amplitude $a_\text{vac}(\Omega)$ of wave pumping interferometer through detector port describe vacuum fluctuations.

Below we assume that mirrors displacements are much smaller than the light wave length so we can expand exponent in series, for example, as $e^{ikx_2}\simeq 1 +ikx_2$ and so on. For displacements we widely use frequency (spectral) domain, for example,
\begin{align*}
x_2(t)=\int_{-\infty}^\infty x_2(\Omega)\, e^{-i\Omega t}\, \frac{d\Omega}{2\pi}
\end{align*}

The symmetry is a crucial feature of this interferometer. Indeed, the waves circulating clockwise and counterclockwise inside the ring cavity contain information about GW action and positions of all mirrors and platform $P$, however, after recombining on the beam splitter the amplitude $d_P(\Omega)$ of output wave in detector port depends only on positions $x_2$ of mirror $M_2$, $x_4$ of mirror $M_4$ and GW's metric perturbation $h$ \cite{2008Kawamura}. Indeed, due to symmetry the clockwise and counterclockwise waves contain {\em identical} information on displacements $y_P,\ y_1, \ y_3$ which cancel after recombination on the beam splitter $M_{bs}$ (actually after subtraction). One can calculate amplitude $d_P(\Omega)$ on detector in spectral domain (details of calculations are presented in Appendix~\ref{app1}):
\begin{align}
\label{d}
d_P(\Omega)&= -ia_\text{vac}(\Omega)\, \frac{\theta^4-R}{1-R\theta^4}+\\
&\quad + \frac{T^2\,A\, ik\,\theta_0\theta\,(\theta_0^2-\theta^2) \big( x_4(\Omega) +
x_2(\Omega) \big)}{\sqrt 2(1-R\theta^4)(1-R\theta_0^4)}+\nonumber\\
&\quad + \frac{4i\,T^2\, A\,\theta_0^2\theta^2\, g(\Omega)}{(1-R\theta^4)(1-R\theta_0^4)},\nonumber\\
\label{g}
g(\Omega)&\equiv h(\Omega)\, kL\, \frac{\sin^2 \frac{\Omega\tau}{2}\, \cos\Omega\tau}{\Omega\tau},\\
\theta_0&=e^{i\delta\tau},\quad \theta= e^{i(\delta+\Omega)\tau},\quad \tau=\frac{L}{c}\, .
\end{align}
Here $R=\sqrt{1-T^2}$ is reflectivity of mirror $M_1$. We assume that laser frequency $\omega$ is detuned by $\delta$ from resonance frequency $\omega_0=\omega-\delta$ of ring interferometer. It is worth repeating that fluctuational amplitude $a(\Omega)$ is absent in $d_P(\Omega)$ --- it means laser noise exclusion.

The analyzed interferometer is a kind of speed meter. Recall that speed meter was proposed in 1990 by V.~Braginsky and F.~Khalili \cite{90a1BrKh} as a QND device \cite{96a1KhLe, 00a1BrGoKhTh, 02a1Pu, 02a1PuCh, 03a1Ch, 04a1Da}, allowing to overcome the Standard Quantum Limit. Note that combination $\theta\theta_0(\theta_0^2-\theta^2) \big( x_4(\Omega) + x_2(\Omega)\big)$ presented in spectral domain in second term (\ref{d}) may be written in time domain as
\begin{align*}
\theta_0^4&\big( x_4(t-\tau) -x_4(t-3\tau)+ x_2(t-\tau)- x_2(t-3\tau)\big)\simeq\\
&\simeq \theta_0^4 2\tau \big(v_2(t-2\tau) + v_4(t-2\tau)\big)
\end{align*}
In last equality we apply the long wave approximation $\Omega\tau\ll 1$ and expand $x_{2,\ 4}(t-\tau)\simeq x_{2,\ 4}(t-2\tau) +\tau v_{2,\ 4}(t-2\tau)$. Actually the detector analyzed gauges not the positions of the mirrors $x_2,\ x_4$ but rather mirror velocities $v_2,\ v_4$. And the signal containing information on velocities is amplified in resonant manner. This is the reason to call this detector a resonant speed meter.

GW (signal) response of our detector (as well as signal response of other variants of speed meter) is also smaller if compared with GW response of conventional Fabry-Perot cavity (analog of LIGO interferometer) by a factor which in spectral domain is equal to (see (\ref{g}))
\begin{align}
\label{decrease}
\frac{4\sin^2 \frac{\Omega\tau}{2}\, \cos\Omega\tau}{\Omega\tau}\simeq \Omega\tau\, .
\end{align}
(In last equality we used long wave approximation.) In time domain and in long wave approximation it means that GW signal is proportional to $\tau \dot h(t)$ instead of $h(t)$ in conventional GW detector.

\section{Displacement noise free speed meter}\label{DFI}

\begin{figure}
\includegraphics[width=0.45\textwidth]{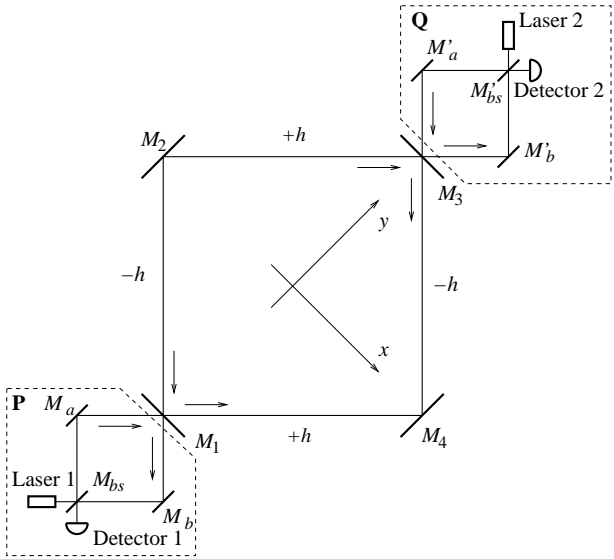}
\caption{Design of double pumped speed meter. Laser 1, detector 1 and auxiliary mirrors $M_{bs},\ M_a,\ M_b$ are rigidly mounted on movable platform $P$. Laser 2, detector 2 and auxiliary mirrors $M'_{bs},\ M_a',\ M'_b$ are rigidly mounted on platform $Q$. Mirror $M_1$ ($M_3$) has finite transmissivity $T$ for light emitted by laser 1 (2) but it is completely reflective for light from laser 2 (1). GW propagates perpendicularly to the plane of interferometer. Recycling cavity is composed of mirrors $M_1,\ M_2,\ M_3,\ M_4$. The light waves pumped by laser 1 (2) and circulating counterclockwise (clockwise) are shown by arrows.}\label{sm2}
\end{figure}

Let us consider the design of a double pumped interferometer shown in Fig.~\ref{sm2}. It only differs from interferometer in Fig.~\ref{sm} by an additional laser 2 pumping interferometer through mirror $M_3$. We assume that laser 2, detector 2 with reflective mirrors $M'_a,\ M'_b$ and beam splitter $M'_{bs}$ are rigidly mounted on platform $Q$, which can move as a free mass along $y$ axis. Power and frequency of laser 2 are the same as those of laser 1. We assume that beams of laser 1 and 2 have orthogonal polarization in order to exclude nonlinear coupling. We also assume that mirror $M_1$ has transmissivity $T$ for waves emitted by laser 1 but it is completely refractive for waves emitted by laser 2. By the same way the mirror $M_3$ has transmissivity $T$ for waves emitted by laser 2 but  is completely refractive for waves emitted by laser 1. (It may be realized using mirrors whose transmissivity depends on polarization of light or, alternatively, the lasers may operate at different optical frequencies and transmissivities of mirrors $M_1,\ M_3$ resonantly depend on frequency). So light emitted by laser 1 completely returns through mirror $M_1$ to platform $P$ and formula (\ref{d}) for output wave $d$ is valid.

The symmetry of interferometer relatively axis $x$ plays a key role. Indeed, the wave from laser 1 circulating counterclockwise in ring cavity and wave from laser 2 circulating clockwise (the directions of their propagation are shown by arrows in Fig.~\ref{sm2}) contain the {\em same} information on positions of mirrors $M_1$ --- $M_4$. The same is valid for clockwise wave from laser 1 and counterclockwise wave from laser 2. Hence, after recombination on beam splitters the wave in detector 1 port (with the amplitude $d_P(\Omega)$) and wave in detector 2 port have to contain {\em identical} information on displacement $x_2$ and $x_4$ (information on positions $y_1, y_3$, as well as on platform positions $y_P \ (y_Q)$, cancels after recombination on beamsplitters --- see formula (\ref{d})).

It should be underscored that the contribution of GW into phase advances of counterclockwise wave from laser 1 and of clockwise wave from laser 2 have opposite signs. Indeed, for counterclockwise wave from laser 1 GW dimensionless metric $h$ has first positive sign (between mirrors $M_1$ and $M_4$) then negative sign and so on, but for clockwise wave from laser 2 GW metric $h$ has first negative sign (between mirrors $M_3$ and $M_4$) then positive sign and so on --- it is obvious from Fig.~\ref{sm2}.

Summing up we can find the amplitude $d_Q(\Omega)$ of wave in detector 2 port just rewriting formula (\ref{d}) for $d_P(\Omega)$ with opposite sign at term proportional to $g$ which describes GW action:
\begin{align}
\label{d2}
d_Q(\Omega)&= -ib_\text{vac}(\Omega)\, \frac{\theta^4-R}{1-R\theta^4}+\\
&\quad + \frac{T^2\,\theta_0\theta\,(\theta_0^2-\theta^2) A\, ik\big( x_4(\Omega) +
x_2(\Omega) \big)}{\sqrt 2(1-R\theta^4)(1-R\theta_0^4)}-\nonumber\\
&\quad - \frac{4i\,T^2\, A\,\theta_0^2\theta^2\, g(\Omega)}{(1-R\theta^4)(1-R\theta_0^4)}\,\nonumber.
\end{align}
Here amplitude $b_\text{vac}(\Omega)$ describes vacuum fluctuation pumped into interferometer through detector 2 port.

Now constructing the following linear combination of both detectors responses
\begin{align}
\label{C}
C&= \frac{d_P(\Omega)-d_Q(\Omega)}{\sqrt 2}=
\frac{a_\text{vac}(\Omega)-b_\text{vac}(\Omega)}{i\sqrt 2}\, \frac{\theta^4-R}{1-R\theta^4}+\\
&\quad + \frac{4\sqrt 2\, i\,T^2\, A\,\theta_0^2\theta^2\, g(\Omega)}{(1-R\theta^4)(1-R\theta_0^4)}\, ,
\nonumber
\end{align}
we are able to cancel the displacement noise of mirrors $M_2,\ M_4$.

Note that the ``price'' for displacement noise exclusion is given by a decrease in response of signal (GW), however, this decrease is the same as for the conventional speed meter. For example, for conventional Fabry-Perot cavity used as GW detector (analog of LIGO interferometer) the signal response is described by a similar formula (\ref{C}) with only one substitution $g_\text{conv}$ instead of $g$, where
\begin{align*}
g_\text{conv} & \equiv h(\Omega)\, kL\, \frac{\sin \Omega\tau}{\Omega\tau},\quad
g_\text{conv}\simeq h(\Omega)\, kL\, ,\quad \text{if}\ \Omega\tau \ll 1
\end{align*}
Recall that in long wave approximation $g\simeq h(\Omega)\, kL\, \Omega\tau$ (see (\ref{decrease})). Therefore, decrease in signal (or signal to noise ratio) for analyzed DFI speed meter as compared with conventional interferometric GW detector is {\em only} by factor about $ \Omega \tau\ll 1$ which much larger than by factor $(\Omega\tau)^3$ for double Mach-Zehnder DFI \cite{2006_interferometers_DNF_GW_detection} or $(\Omega\tau)^2$ for two double pumped Fabry-Perot cavities \cite{08a1RaVy}.

Note that resonance gain is presented both in our DFI speed meter (final formula (\ref{C}) demonstrates it) and in conventional Fabry-Perot GW detector. However, the resonance gain is almost compensated by small factor $\Omega\tau$. In order to show it we rewrite formula (\ref{C}) in long wave approximation  expanding in series $\theta\simeq 1+i(\Omega+\delta)\tau,\ \theta_0\simeq 1+i\delta \tau$, $R\simeq 1-T^2/2$ (due to $\Omega\tau\ll 1$ and $T\ll 1$):
\begin{align}
\label{Clw}
C_{\Omega\tau\ll 1}& \simeq
\frac{a_\text{vac}(\Omega)-b_\text{vac}(\Omega)}{i\sqrt 2}\,
\frac{\gamma+i(\delta+\Omega)}{\gamma-i(\delta+\Omega)}+\\
& + 2\sqrt 2\, i\,A\,kL h(\Omega)\left(\frac{\gamma\Omega}{
\big(\gamma-i(\delta+\Omega)\big)\big(\gamma-i\delta\big)}\right) ,
\nonumber
\end{align}
where $\gamma=T^2/8\tau$ is relaxation rate (half bandwidth) of ring cavity. Obviously, the absolute value of fraction inside large round brackets has maximum at zero detuning ($\delta=0$) and it does not exceed the unity:
\begin{align}
\big|C_{\Omega\tau\ll 1}\big|& < \left|
\frac{a_\text{vac}(\Omega)-b_\text{vac}(\Omega)}{i\sqrt 2}\right|+
2\sqrt 2\, \,A\,kL h(\Omega)
\end{align}
Hence, the signal response is practically the same as for simplest single round trip GW detector \cite{2008_Tarabrin}.

\section{Conclusion}\label{conclusion}

In this paper we have analyzed the operation of double pumped resonant speed meter, performing the displacement noise free gravitational-wave detection. We have demonstrated that it is possible to produce a linear combination of two response signals which cancels the displacement fluctuations of the mirrors. At low frequencies GW response in our DFI turns out to be better (sensitivity decreases by factor $\Omega\tau$ only) than that in Mach-Zehnder-based DFIs \cite{2006_interferometers_DNF_GW_detection} or two double pumped Fabry-Perot cavities \cite{08a1RaVy}.

It is worth noting that the symmetry plays a key role in analyzed DFI. First, due to the symmetry relatively clockwise and counterclockwise waves we have the opportunity of excluding both laser noise and noise of displacements $y_P\ (y_Q),\ y_1,\ y_3$ in the output signal in each detector port. Second, the symmetry relatively axis $x$ allows excluding information on displacements $x_2,\ x_4$ and converting resonant speed meter into DFI.

Our analysis is based on the statement that we can work in laboratory frame using TT gauge if laser and detector (measuring laser light wave after its reflection from the ring cavity having movable mirrors) are mounted on the same platform. However, this statement was proved for a single round trip configuration \cite{2008_Tarabrin} and, strictly speaking, it should be checked  independently for configuration of displacement noise free resonant speed meter analyzed in this article.

The proposed configuration of DFI is a gedanken (mental) device, however, it may be a promising base candidate for the future configurations of GW detectors with displacement and laser noise exclusion which, in turn, will allow overcoming the Standard Quantum Limit.

\acknowledgments
We are grateful to V.B.~Braginsky, S.L.~Danilishin, F.Ya.~Khalili, A.~Nishizawa, S.P.~Tarabrin for fruitful discussions, especially, to A.~Nishizawa for paying attention on article \cite{2008Kawamura} and making critical remarks. This work was supported by LIGO team from Caltech, in part by NSF and Caltech grant PHY-0651036 and by Grant of the President of the Russian Federation NS-5178.2006.2.

\appendix

\section{Single pumped Sagnac interferometer}\label{app1}

In this Appendix we present derivation of formula (\ref{d}) for output of resonant speed meter based on interferometer proposed in \cite{2008Kawamura}.

\begin{figure}
\includegraphics[width=0.45\textwidth]{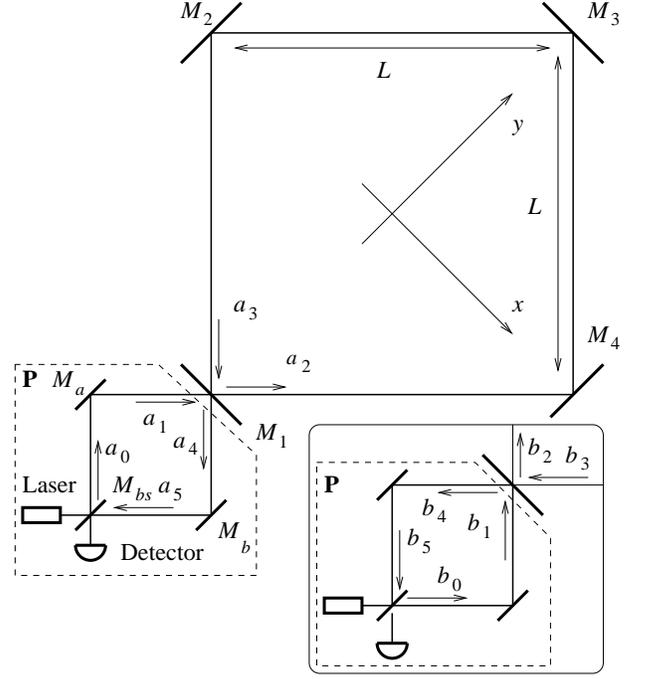}
\caption{Simplified model of speed meter as shown in Fig.~\ref{sd}. Light waves circulating counterclockwise are shown by arrows, waves circulating clockwise are shown on insert.}\label{sm}
\end{figure}

First we calculate mean amplitudes of wave propagating counterclockwise, starting from conditions on input $M_1$ mirror (see notations in Fig.~\ref{sm}):
\begin{align}
A_2&= iTA_1 -R A_3,\quad A_3 =-\theta_0^4 A_2,\nonumber\\
&\Rightarrow\quad A_2=\frac{iTA_1}{1-R\theta_0^4},\\
A_4 &= iT\, A_3 - RA_1,\quad \Rightarrow \ A_4=\frac{\theta_0^4-R}{1-R\theta_0^4}\, A_1 ,\\
A_1&= -\frac{A}{\sqrt 2}\, e^{i\phi}, \quad
A_5 =\frac{\theta_0^4-R}{1-R\theta_0^4}\, \frac{-Ae^{2i\phi}}{\sqrt 2}
\end{align}
Here, $A$ is amplitude of wave emitting by laser, $A_1$ is amplitude of wave falling on mirror $M_1$, $\phi$ is the phase advance of wave propagating from beam splitter to mirror $M_1$ (we assume it as a constant).

By similar way we calculate mean amplitudes for wave propagating clockwise:
\begin{align}
B_2 &=\frac{iTB_1}{1-R\theta_0^4},\quad
B_4=\frac{\theta_0^4-R}{1-R\theta_0^4}\, B_1 ,\\
B_1&= \frac{iA}{\sqrt 2}\, e^{i\phi}, \quad
B_5 =\frac{\theta_0^4-R}{1-R\theta_0^4}\, \frac{iAe^{2i\phi}}{\sqrt 2}
\end{align}
Now we can calculate output amplitude $A_\text{laser}$ in laser port and $A_\text{det}$ in detector port:
\begin{align}
A_\text{laser}& = \frac{-B_5+iA_5}{\sqrt 2} =\frac{\theta_0^4-R}{1-R\theta_0^4}\big(-iA)e^{2i\phi},\\
A_\text{det} &= \frac{iB_5-A_5}{\sqrt 2} = 0
\end{align}
Below we put $\phi=0$.

Now we calculate small amplitude. We start from wave propagating counterclockwise inside ring cavity. We have for amplitudes $a_2,\ a_3$ of wave circulating inside ring cavity (recall that after each reflection from mirror amplitude reverses its sign):
\begin{align}
a_3 &= -\theta^4 a_2 -\theta_0\theta^3(-1)^2\, u_4 + \theta_0^2\theta^2(-1)\, u_3 +
\theta_0^3\theta\, u_2+u_g\nonumber\\
u_4 &\equiv A_2\, \sqrt 2\,ikx_4,\quad u_3\equiv A_2\, \sqrt 2\,iky_3,\\
u_2 & \equiv A_2\, \sqrt 2\,ikx_2,\quad u_1 \equiv A_2\, \sqrt 2\, iky_1,\\
u_g &\equiv A_2\theta_0 j\theta^3+(-A_2)\theta_0^2j\theta^2+A_2\theta_0^3j\theta-
A_2\theta_0^4j\nonumber\\
j&e^{-i\Omega t}\equiv \frac{h\omega}{2}\int_{t-\tau}^te^{-i\Omega t'}\, dt',\quad
j=\frac{h\omega\tau}{2}\frac{(1-e^{i\Omega\tau})}{-i\Omega\tau},\nonumber\\
u_g&=A_2\theta_0^2\theta^2\, 4ih\omega\tau\, \frac{\sin^2 \frac{\Omega\tau}{2}\, \cos\Omega\tau}{\Omega\tau}\\
a_2 &= iTa_1-Ra_3 - R\theta_0^4 u_1,\\
a_2 &=\frac{iT a_1}{1-R\theta^4} +
\frac{R\big(\theta_0\theta^3\, u_4 + \theta_0^2\theta^2\, u_3-
\theta_0^3\theta\, u_2- \theta_0^4 u_1-u_g\big)}{1-R\theta^4},\nonumber\\
a_3 &= \frac{-iT\theta^4\, a_1}{1-R\theta^4}-\\
&\quad - \frac{\theta_0\theta^3\, u_4 + \theta_0^2\theta^2\, u_3-
\theta_0^3\theta\, u_2 - R\theta^4\theta_0^4 u_1- u_g}{1-R\theta^4}\nonumber
\end{align}
Now we can find amplitude $a_4$ of wave reflected from ring cavity
\begin{align}
 a_4 &= iT\, a_3-Ra_1 + (-R A_1) \sqrt 2\, iky_1,\quad A_1=\frac{1-R\theta_0^4}{iT}\, A_2,\nonumber\\
 a_4 &= \frac{\theta^4-R}{1-R\theta^4}\, a_1 +\nonumber\\
   &\quad + iT\frac{-\theta_0\theta^3\, u_4 - \theta_0^2\theta^2\, u_3+
        \theta_0^3\theta\, u_2 + R\theta^4\theta_0^4 u_1 + u_g}{1-R\theta^4}-\nonumber\\
    &\qquad -R\frac{1-R\theta_0^4}{iT}\, u_1=\nonumber\\
 &= \frac{\theta^4-R}{1-R\theta^4}\, a_1 +
    iT\frac{-\theta_0\theta^3\, u_4 - \theta_0^2\theta^2\, u_3+
        \theta_0^3\theta\, u_2 + u_g}{1-R\theta^4}-\nonumber\\
\label{a4}        
   &\qquad -\frac{R\big(1+\theta^2\theta_0^4-R(\theta^4+\theta_0^4)\big)}{iT\big(1-R\theta^4\big)}\, u_1
\end{align}

Above $a_1$ and $a_4$ (as well as $a_2,\ a_3$) are amplitudes on mirror $M_1$. Now we should take into account displacement $y_p$ of platform $P$. First, we recalculate formula (\ref{a4}) through input fluctuational amplitude $a$ from laser and amplitude $a_\text{vac}$ from detector port:
\begin{align}
a_1 &= \frac{-a +ia_\text{vac}}{\sqrt 2} + A_1\, \sqrt 2 iky_p\, ,
\end{align}
Second, we find the amplitude $a_5$ of counterclockwise wave on beam splitter $M_{bs}$:
\begin{align}
a_5 = a_4 +A_4\, \sqrt 2 iky_p\, .
\end{align}
Finally we get
\begin{align}
\label{a5}
a_5&= \frac{\theta^4-R}{1-R\theta^4}\, \left[\frac{-a +ia_\text{vac}}{\sqrt 2} \right]
-\\
&\quad - iT\frac{\theta_0\theta^3\, u_4 + \theta_0^2\theta^2\, u_3-
\theta_0^3\theta\, u_2 -u_g}{1-R\theta^4}-\nonumber\\
&\quad -\frac{R\big(1+\theta^2\theta_0^4-R(\theta^4+\theta_0^4)\big)}{iT\big(1-R\theta^4\big)}\, u_1
+ \nonumber\\
&\quad +\frac{-A}{\sqrt 2}\left(\frac{\theta^4-R}{1-R\theta^4}+\frac{\theta_0^4-R}{1-R\theta_0^4} \right) \sqrt 2 iky_p
\, . \nonumber
\end{align}

For wave circulating counterclockwise we can rewrite the formulas above for waves circulating counterclockwise. For amplitude $b_4$ of wave leaving ring cavity we rewrite formula (\ref{a4}) using substitutions $x_2 \to x_4,\ x_4\to x_2,\ A_2\to B_2,\ u_g\to v_g$:
\begin{align}
b_4 &= \frac{\theta^4-R}{1-R\theta^4}\, b_1 -
iT\frac{\theta_0\theta^3\, v_2 + \theta_0^2\theta^2\, v_3-
\theta_0^3\theta\, v_4 -v_g}{1-R\theta^4}-\nonumber\\
\label{b4}
&\quad -\frac{R\big(1+\theta^2\theta_0^4-R(\theta^4+\theta_0^4)\big)}{iT\big(1-R\theta^4\big)}\, v_1,
\\
v_1 &= B_2\, \sqrt 2\,ikxy_1,\quad v_2=B_2\, \sqrt 2\,ikx_2,\\
v_3 &=B_2\, \sqrt 2\,iky_3,\quad v_4=B_2\, 2ikx_4,\quad v_k=-iu_k,\\
v_g&=-B_2\theta_0^2\theta^2\, 4ih\omega\tau\, \frac{\sin^2 \frac{\Omega\tau}{2}\,
\cos\Omega\tau}{\Omega\tau}
\end{align}
Comparing formulas for $A_2$ and $B_2$ we see that $v_k=-iu_k$, but $v_g=iu_g$.
And for amplitude $b_5$ on beam splitter we have
\begin{align}
\label{b5}
b_5 &= \frac{\theta^4-R}{1-R\theta^4}\, \left[\frac{ia - a_\text{vac}}{\sqrt 2} \right] -\\
&\quad - iT\frac{\theta_0\theta^3\, v_2 + \theta_0^2\theta^2\, v_3-
\theta_0^3\theta\, v_4 -v_g}{1-R\theta^4}-\nonumber\\
&\quad -\frac{R\big(1+\theta^2\theta_0^4-R(\theta^4+\theta_0^4)\big)}{iT\big(1-R\theta^4\big)}\, v_1
+ \nonumber\\
&\quad +\frac{iA}{\sqrt2}
    \left(\frac{\theta^4-R}{1-R\theta^4}+\frac{\theta_0^4-R}{1-R\theta_0^4}\right) 
    \sqrt 2 ik y_p \, . \nonumber
\end{align}

Now we can calculate amplitude $d$ of output wave in detector port. Due to relations $v_k=-iu_k$ and $v_g=iu_g$ only terms proportional to $u_2,\ u_4,\ v_2,\ v_4$ and $u_g,\ v_g$ ``survive'':
\begin{align}
d_P &= \frac{ib_5-a_5}{\sqrt 2}=
\frac{\theta^4-R}{1-R\theta^4}\,(-i)a_\text{vac}+\\
&\quad +
\frac{iT\big[\theta_0\theta\, (u_4 +u_2)(\theta^2-\theta_0^2)-
2u_g\big]}{\sqrt 2(1-R\theta^4)}\nonumber
\end{align}
From this formula one can easy obtain formula (\ref{d}) in Sec.~\ref{simple}.

\end{document}